\documentclass[doublecol]{epl2} 

\newcommand{\bra}{\left\langle}
\newcommand{\ket}{\right\rangle}

\newcommand{\vecr}{{\boldsymbol r}}
\newcommand{\vecv}{{\boldsymbol v}}

\newcommand{\veclam}{{\boldsymbol \lambda}}

\title{Rocking ratchet based on F$_1$-ATPase in the absence of ATP}

\author{Kumiko Hayashi, Hisatsugu Yamasaki and Mitsunori Takano}
\shortauthor{K. Hayashi \etal}

\institute{Department of Physics, Waseda university - 3-4-1 Okubo, 
Shinjuku, Tokyo 169-8555, Japan.}

\abstract{
Bartussek, H\"anggi and Kissner studied a rocking ratchet system, in which  
a Brownian particle is subject to an asymmetric periodic potential together 
with an oscillating force,  and found that the direction of the macroscopic 
current can be reversed by changing the parameter values characterizing 
the model [{\it Europhys. Lett.}, {\bf 28} (1994) 459]. In this letter, we 
apply their ratchet theory to a rotary motor-protein, F$_1$-ATPase. In this 
work,  we construct a model of a rocking ratchet in which F$_1$-ATPase 
rotates not as a result of ATP hydrolysis but through the influence of 
an oscillating force. We then study the motion of F$_1$-ATPase on the 
basis of molecular dynamics simulations of this coarse-grained protein 
model. Although in the absence of ATP, F$_1$-ATPase exhibits directionless 
Brownian motion when there exists no oscillating force, we observe 
directional motion when we do apply an oscillating force. Furthermore, 
we observe that the direction of rotation is reversed when we change 
the oscillation frequency.}

\begin{document}

\maketitle

\section{Introduction}
Recently, theories of non-equilibrium statistical mechanics have been 
applied to many biological systems. Among such studies, the 
fluctuation-dissipation theorem has been investigated with the purpose of 
understanding the motion of myosins in an actin network \cite{mizuno}, 
the fluctuation theorem has been used to quantify the fluctuations of cell 
motion \cite{hayashi1}, and the Jarzynski equality and the Crooks theorem 
have been applied to RNA hairpins in order to measure their free energies 
\cite{ritort1}. Recent technical developments have made it possible to  
investigate small systems, including biological systems, and this has 
provided another way to test theories of non-equilibrium statistical 
mechanics in addition to the investigation of theoretical models 
of stochastic processes. 

Ratchet theories have been proposed to describe certain types of 
non-equilibrium statistical behavior \cite{prost,hanggi1,hanggi2}, and 
they predict that together, spatial asymmetry and non-equilibrium effect, 
can cause directional motion in stochastic systems \cite{hanggi1}. Among the 
various ratchet models, that consisting of a Brownian particle in an 
asymmetric periodic potential together with an oscillating force is called 
a ``rocking ratchet.'' For such ratchets, it has been shown that the 
direction of the current can be reversed by changing the parameter values 
characterizing the model \cite{hanggi2}. Although the application of such 
current reversal to nano-sized systems is mentioned in Ref. \cite{hanggi2}, 
as far as we know, there is yet no such experiment.  In this letter, we 
numerically investigate the rocking ratchet theory in application to a 
nano-sized  bio-molecule to which many theories of non-equilibrium 
statistical mechanics have been applied in recent years. 

We study a rotary motor-protein, F$_1$-ATPase, in which a $\gamma$-subunit 
rotates in the center of  $\alpha_3\beta_3$-subunits \cite{noji,toshio}. 
In this letter, we construct a model of a rocking ratchet in which 
F$_1$-ATPase rotates through the influence of an oscillating force, and 
particularly study the motion of F$_1$-ATPase in the absence of ATP.  
Because in this case no nucleotide is attached to the $\beta$-subunits, 
such a state is called a nucleotide-free state.  In the absence of ATP, 
it was observed that the $\gamma$-subunit exhibits directionless Brownian 
motion  with rotational steps of $\pm 120^\circ$ \cite{fujisawa}. This 
indicates that the rotational potential experienced by a $\gamma$-subunit 
is periodic with a period of $120^\circ$, reflecting the three-fold symmetry 
of $\alpha_3\beta_3$-subunits. Then, if this periodic potential, which 
reflects the interaction between  a $\gamma$-subunit and 
$\alpha_3\beta_3$-subunits, is asymmetric, we expect that a $\gamma$-subunit 
will display a directional rotation when we apply an oscillating force.

In order to test this conjecture, we performed molecular dynamics simulations 
of a protein model, in which a protein molecule is regarded as consisting 
of $\alpha$-carbon atoms, employing a coarse-grained representation of the 
amino-acid residues \cite{hayataka}.  We find that the $\gamma$-subunit 
rotates counterclockwise  when viewed from the membrane side in the case 
that the oscillation period is larger than the characteristic time scale 
of the system dynamics, and it rotates clockwise in the opposite case.

\section{ F$_1$-ATPase in the presence of ATP}
F$_1$-ATPase is a rotary motor-protein in which a $\gamma$-subunit rotates 
in the center of $\alpha_3\beta_3$-subunits due to ATP hydrolysis 
\cite{noji,toshio}.  In the presence of ATP, a $\gamma$-subunit rotates 
counterclockwise  when viewed from the membrane side and  makes a 
$120^\circ$ step as a result of the consumption of  one ATP molecule. 
(In this letter, the counterclockwise direction is regarded as the plus 
direction.) Each step of $120^\circ$ is divided into $40^\circ$  and 
$80^\circ$ substeps, caused by the two chemical reactions involved in this 
process, i.e. the product release and the ATP binding, respectively.  


%
\begin{figure}
\begin{center}
\includegraphics[width=8cm]{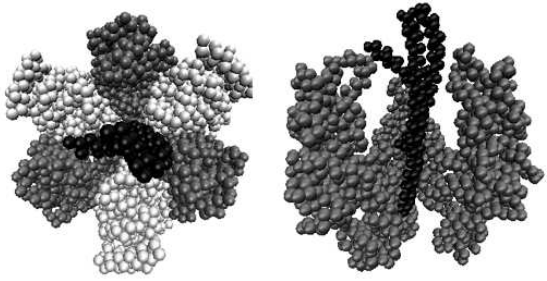} 
\end{center}
\caption{Schematics of the F$_1$-ATPase in a nucleotide-free state,  
where the F$_1$-ATPase consists of the $\alpha$-subunits (light gray), 
the $\beta$-subunits (gray) and the $\gamma$-subunit (black).  
 $\{ \alpha_{\rm E}\beta_{\rm E}, \alpha_{\rm E}\beta_{\rm E}, \alpha_{\rm E}
\beta_{\rm E} \}$ is composed of the $\alpha_{\rm E}\beta_{\rm E}$-subunits 
of the crystal structure 1BMF \cite{walker}. 
}  
\label{fig:b0}
\end{figure}

X-ray studies of the crystal structure of F$_1$-ATPase reveal that a 
$\beta$-subunit possesses several different structures \cite{walker}.  
The ATP-bound $\beta$-subunit, the ADP-bound $\beta$-subunit and the 
nucleotide-free $\beta$-subunit are denoted by $\beta_{\rm TP}$, 
$\beta_{\rm DP}$ and  $\beta_{\rm E}$, respectively.  While $\beta_{\rm TP}$ 
has a closed form at its C terminus, $\beta_{\rm E}$ has an open form.  
Because the structures of the $\beta$-subunits change as the chemical 
reactions involved in ATP hydrolysis proceed, it is widely believed that 
the coordinated push-pull motion of the C  termini of $\beta$-subunits 
causes the rotation of the $\gamma$-subunit \cite{wang}. Such substeps 
have been observed in numerical studies of coarse-grained protein  models 
by changing the reference structures of the  $\beta$-subunits so as to 
create the push-pull motion \cite{koga,pu}. In this letter, although we 
do not study such push-pull  motion, we do  study rotation in  
F$_1$-ATPase.  This rotation is realized in the absence of ATP through 
use of an oscillating force in a coarse-grained protein model (see the  
Appendix). 

\begin{figure}
\begin{center}
\includegraphics[width=8cm]{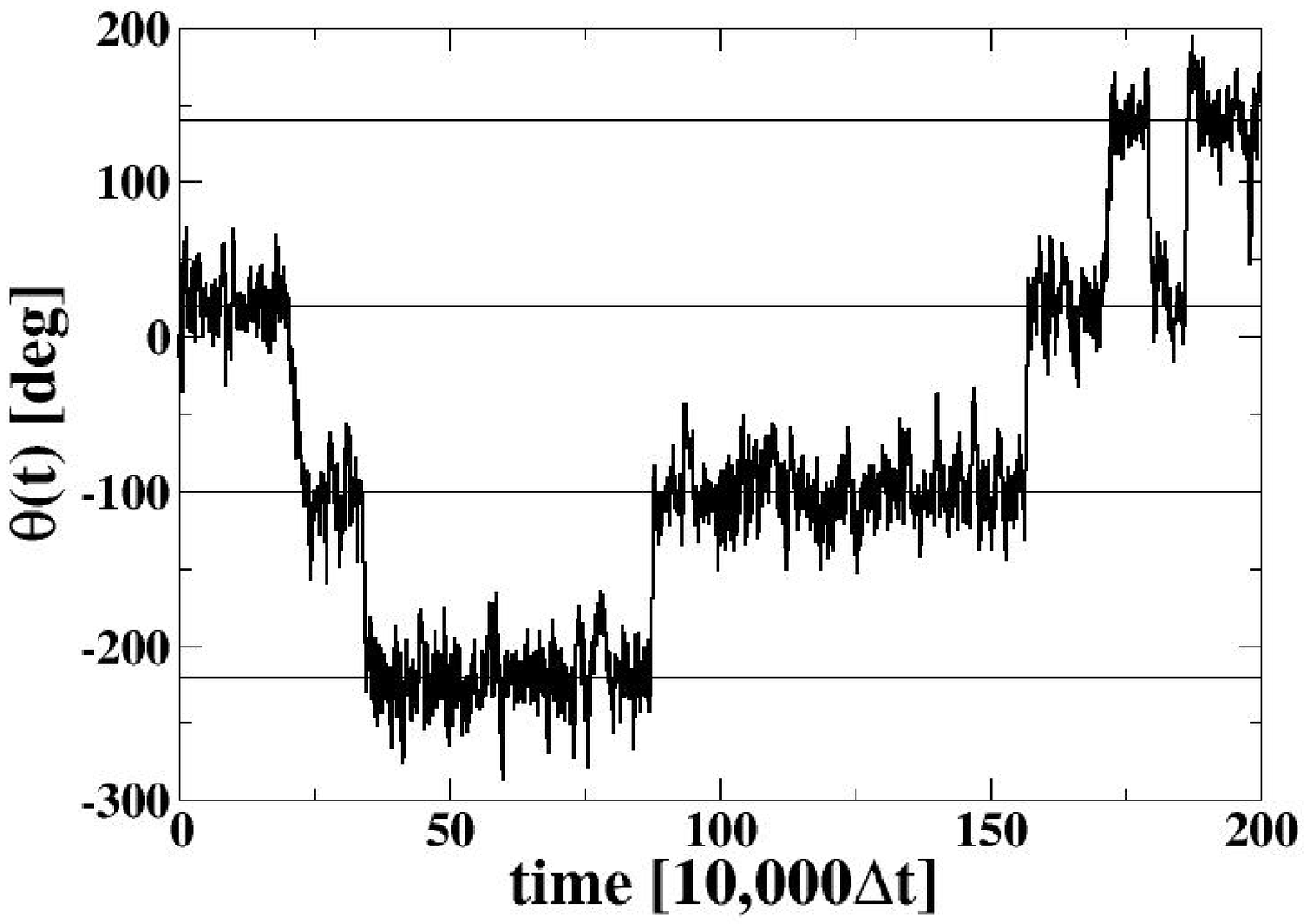} 
\includegraphics[width=8cm]{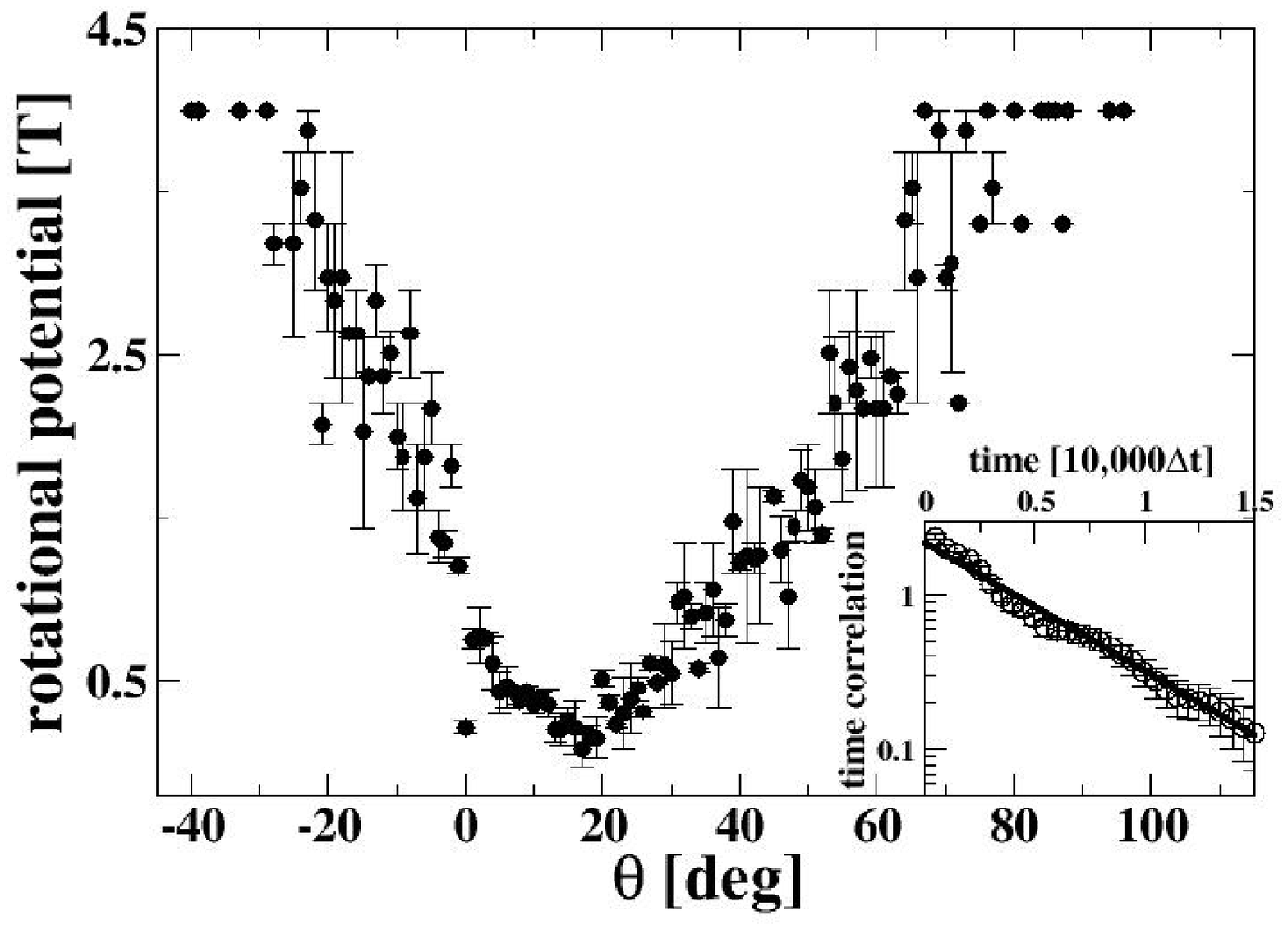}
\end{center}
\caption{(Top) An example trajectory of the angle, $\theta$, of the 
$\gamma$-subunit in a nucleotide-free state. $\theta=0^\circ$ represents 
the position of the $\gamma$-subunit in the crystal structure 1BMF 
\cite{walker}.  The dotted lines represent $\theta=20^\circ+120^{\circ}n$ 
($n=0,\pm 1,\pm 2\cdots$). (Bottom) The rotational potential,  
$-\log(P(\theta))$. (Inset) The time correlation function,  
$\bra \theta(t)\theta(0)\ket - \bra \theta(0) \ket^2$, in dwell time 
(circles). The fitting curve (a thick black curve) represents 
$2.2\exp(-t/5,236\Delta t)$.
}  
\label{fig:b1}
\end{figure}

\section{F$_1$-ATPase in the absence of ATP} 
While a $\gamma$-subunit always makes steps of $+ 120^\circ$ in the presence 
of ATP, both forward, $+ 120^\circ$, and backward, $- 120^\circ$, steps can 
be observed experimentally in the absence of ATP \cite{fujisawa}.  Because 
the structure of $\beta_{\rm E}$, which has an open form, makes the height 
of the rotational potential smaller \cite{fujisawa}, both $+ 120^\circ$ 
and $- 120^\circ$ steps can be observed.  

First, we investigated the Brownian motion of F$_1$-ATPase in equilibrium  
 (without an oscillating force) through numerical simulations. Here, we 
assume that the reference structure of the  $\alpha_3\beta_3$-subunits in  
a nucleotide-free state can be represented by   
$\{ \alpha_{\rm E}\beta_{\rm E}, \alpha_{\rm E}\beta_{\rm E},  
\alpha_{\rm E}\beta_{\rm E} \}$, and that in equilibrium  there is no  
structural change of the $\beta$-subunits. We computed the time evolution 
of the rotation angle $\theta(t)$, which represents the direction of the 
$\alpha$-carbon atoms in the $\gamma$-subunit encircled in the top-left 
region of Fig. \ref{fig:rot}.  (See the Appendix and Ref. \cite{koga} for a 
precise definition.)   In the upper graphs of Fig. \ref{fig:b1}, we plot  
a typical trajectory of $\theta(t)$ in the nucleotide-free state as a 
function of time. Directionless Brownian motion is observed, as there are 
both steps of $+120^\circ$ and $-120^\circ$.  In the upper graphs of 
Fig. \ref{fig:b1},  it is seen that the $\gamma$-subunit stops at angles 
of approximately $20^\circ+120^\circ n$ ($n=0,\pm 1,\pm2,\cdots$). We note 
that it has also been observed experimentally that in the absence of ATP, 
the $\gamma$-subunit stops at angles  shifted by $-20^\circ$ from the 
ATP binding angle \cite{fujisawa}.  Because the ATP binding angle is 
considered to be  $\theta=30^\circ$-$40^\circ$ in our model \cite{koga}, 
the stopping angles plotted in the upper graphs of Fig. \ref{fig:b1} seem 
to correspond to the experimental results \cite{fujisawa}. 

In order to investigate the system dynamics in the intra-well of the 
rotational potential, we analyzed trajectories of $\theta(t)$  that is 
fluctuating around an angle of $20^\circ+120^{\circ}n$ ($n=0,\pm 1,\pm 
2\cdots$). By using the probability distribution, $P(\theta)$,  we compute 
the rotational potential $-\log(P(\theta))$. Then, its asymmetric profile 
is seen in the lower graphs of  Fig. \ref{fig:b1}. In the inset of the 
lower graphs of Fig. \ref{fig:b1}, we further plot the time correlation 
function, $\bra \theta(t)\theta(0)\ket-\bra \theta(0)\ket^2$. From the 
fitting curve, the intra-well relaxation time is estimated to be 
$5,236\Delta t$.  

\begin{figure}
\begin{center}
\includegraphics[width=8cm]{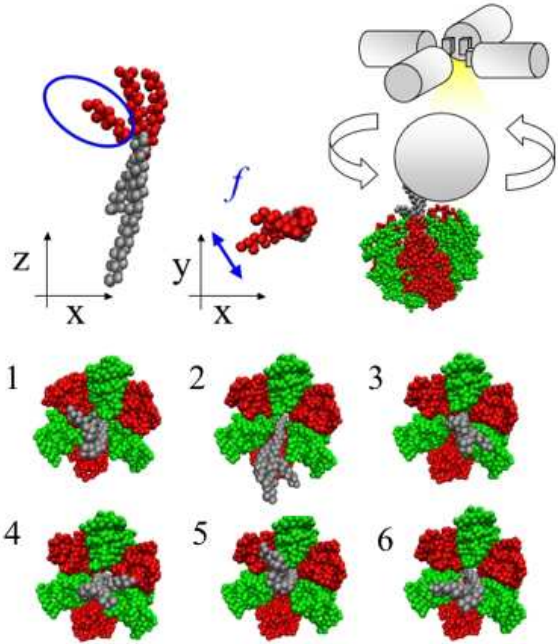}
\end{center}
\caption{(Top) Schematics of the rocking ratchet based on F$_1$-ATPase 
in a nucleotide-free state. (Bottom) In the case that there exists an  
oscillating force with $\tau_\Omega=10,000\Delta t$ and $A=0.005$, the 
$\gamma$-subunit exhibits counterclockwise rotation.  The snapshots 
were taken using VMD \cite{vmd}. 
}
\label{fig:rot}
\end{figure}

\section{Rocking ratchet} 
\begin{figure}
\begin{center}
\includegraphics[width=8cm]{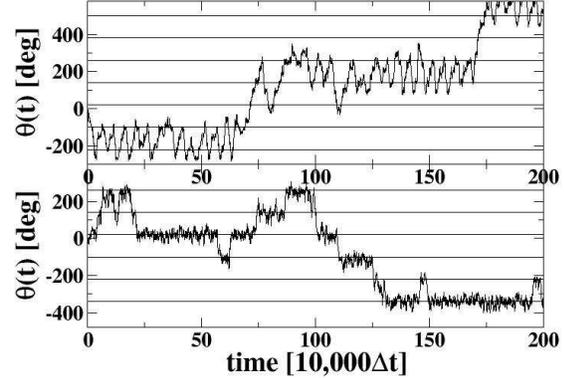}
\end{center}
\caption{Example trajectories in the case of an oscillating force with 
the cases $\tau_{\Omega}=50,000\Delta t$ (Top) and $\tau_\Omega=4,000\Delta 
t$ (Bottom) for $A=0.005$.  The dotted  lines represent 
$\theta=20^\circ+120^{\circ}n$ ($n=0,\pm 1,\pm 2\cdots$). 
}
\label{fig:b3}
\end{figure}

Having verified that our protein model appears to be capable of describing 
the motion of F$_1$-ATPase in the absence of ATP, we next investigated the 
case in which an oscillating force is applied to the $\gamma$-subunit in 
order to create a rocking ratchet.  This force mimics one that can be 
applied by a magnetic tweezers (Fig. \ref{fig:rot}).  Explicitly, we 
apply the oscillating force 
\begin{equation}
(f_x, f_y)=(A\sin2\pi t/\tau_\Omega \sin
\theta, -A\sin 2\pi t/\tau_\Omega \cos\theta) 
\end{equation}
to the $\alpha$-carbon atoms in the $\gamma$-subunit that are not inside  
the $\alpha_3\beta_3$-subunits (marked in red  in Fig. \ref{fig:rot}).  
Note that although in real experiments an oscillating force should be 
applied to the bead to which a $\gamma$-subunit is attached, for 
simplicity, here we apply this force directly to the $\alpha$-carbon 
atoms in the  $\gamma$-subunit. In the lower graphs of Fig. \ref{fig:b3}, 
we plot typical trajectories in the case with an oscillating force of 
periods $\tau_\Omega=50,000\Delta t$ and $\tau_\Omega=4,000\Delta t$, 
respectively.  In Fig. \ref{fig:b3}, it is seen that when  $A=0.005$, 
the range of oscillation of the $\gamma$-subunit is approximately 
$120^\circ$  which is the period of the potential,  and that  the 
$\gamma$-subunit is likely to stop at the same angles as in equilibrium 
even in the case with the oscillating force.

\begin{figure}
\begin{center}
\includegraphics[width=8cm]{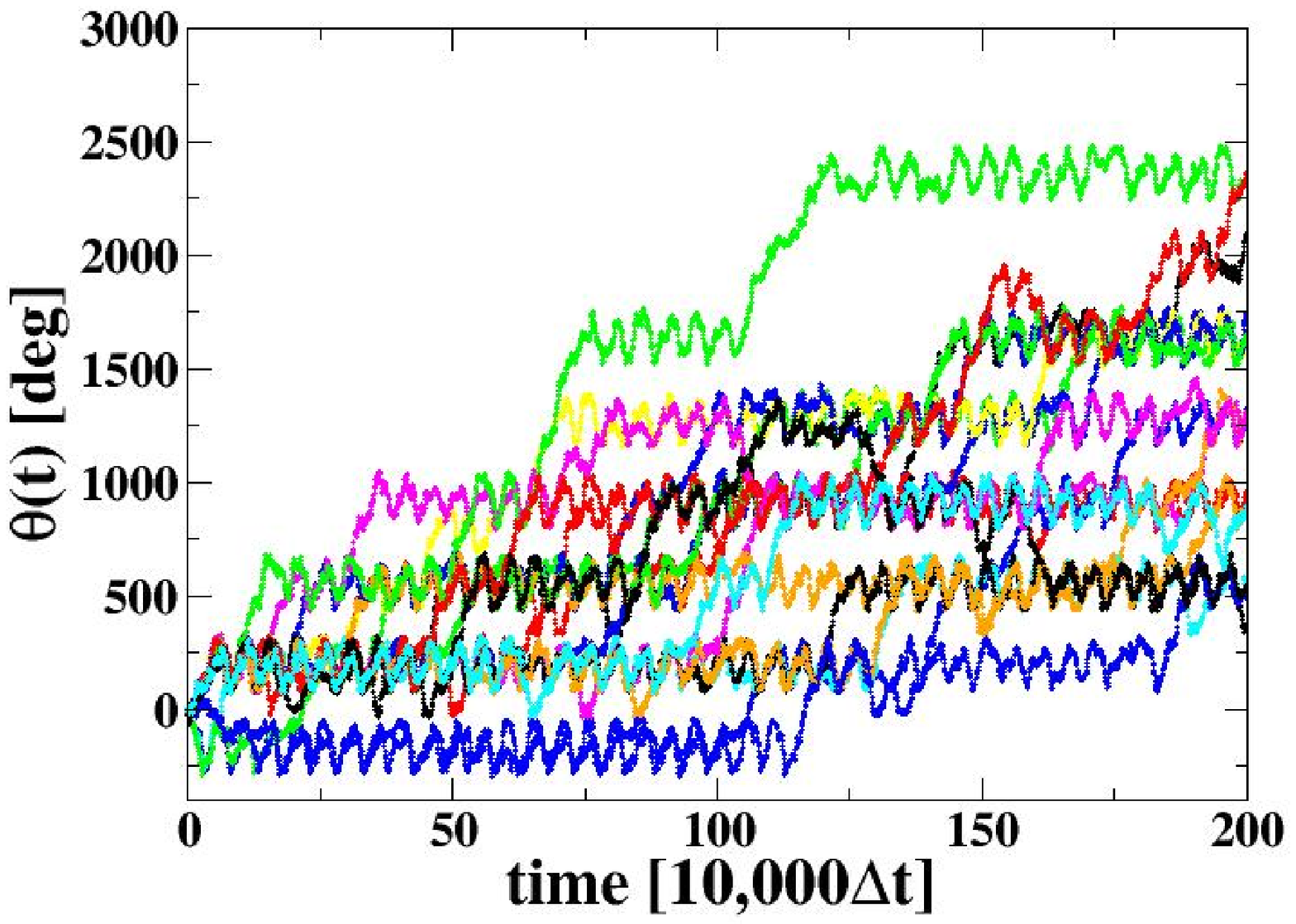} 
\includegraphics[width=8cm]{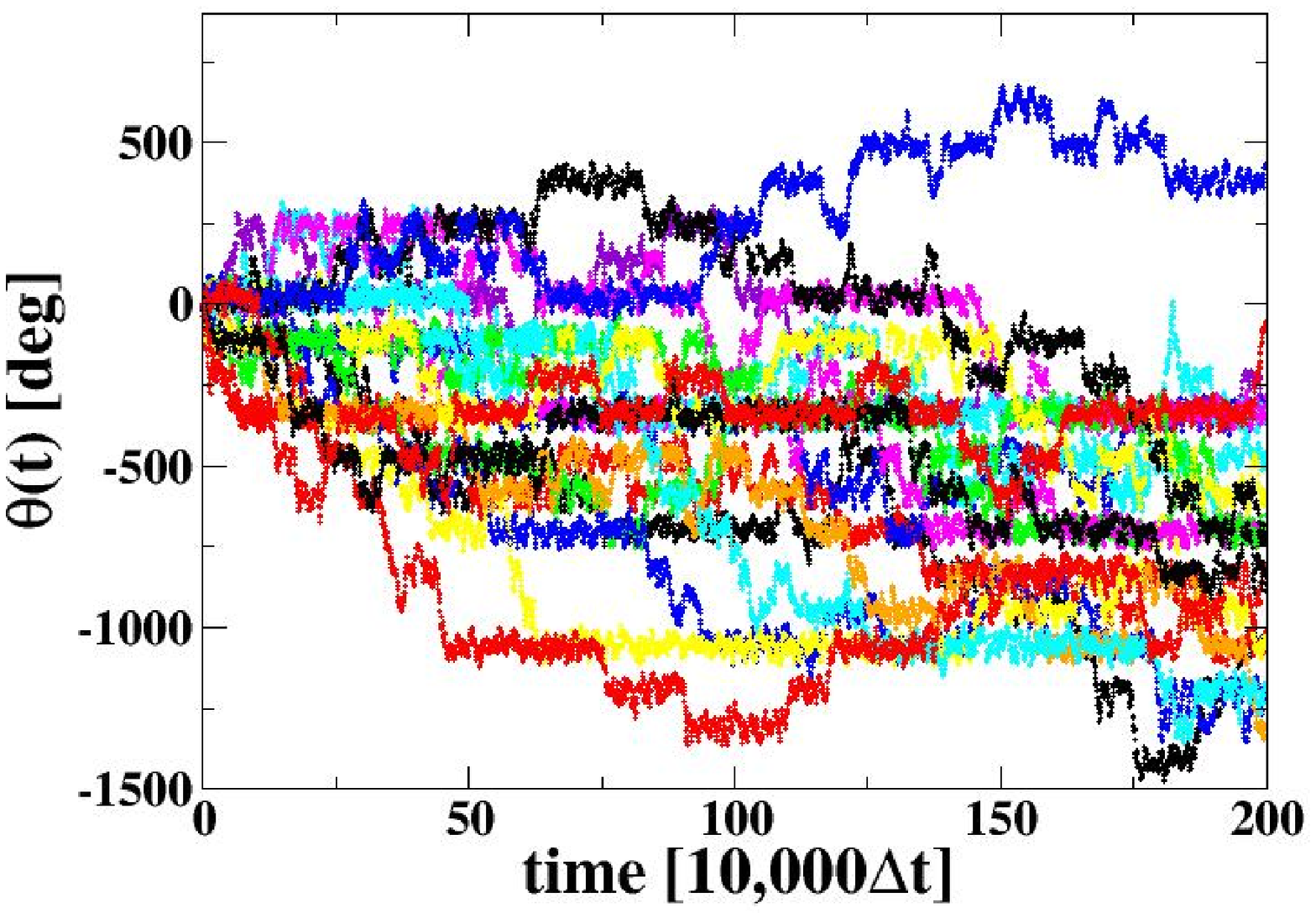}
\end{center}
\caption{Typical trajectories in the cases in which the oscillating 
force has periods, $\tau_{\Omega}=50,000\Delta t$  (Top) and $\tau_
\Omega=4,000\Delta t$ (Bottom) for $A=0.005$.}
\label{fig:b4}
\end{figure}

In Fig. \ref{fig:b4},  we plot $20$ trajectories  in the cases 
$\tau_\Omega=50,000\Delta t$ and $\tau_\Omega=4,000\Delta t$.  There, we 
can see clearly that rotation has a directional preference.  
Interestingly the average direction of rotation in the case with 
$\tau_\Omega=50,000\Delta t$  is opposite to that  in the case  
$\tau_\Omega=4,000\Delta t$. It is thus seen that current reversal occurs 
when we change the oscillation period.  Comparing these results with the 
directions of rotation found in Ref. \cite{hanggi2}, they lead us to 
conjecture that the slope of the potential in the system studied here, 
like that studied there, is somewhat smaller in the counterclockwise 
direction (the plus direction) than that in the clockwise direction. 
Indeed, the rotational potential plotted in the lower graphs of Fig. 
\ref{fig:b1} is consistent with this conjecture.

\begin{figure}
\begin{center}
\includegraphics[width=8cm]{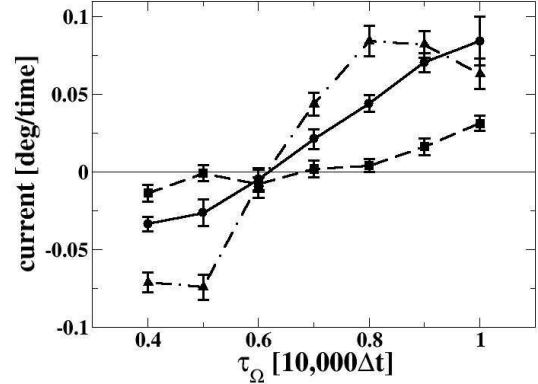}
\end{center}
\caption{The average currents as functions of the oscillation period,   
$\tau_\Omega$, in the cases $A=0.0025$ (squares),  $A=0.005$ (circles) 
and $A=0.0075$ (triangles). 
}
\label{fig:b5}
\end{figure}

In order to determine the value of $\tau_\Omega^*$ at which the current 
reversal occurs, in Fig. \ref{fig:b5}, we plot the average currents 
$\bra \theta(\tau_f)/\tau_f\ket$ as functions of $\tau_\Omega$ for the 
several values of the oscillation amplitude. (Here, $\tau_f=2\times 
10^6\Delta t$.) We find that the current vanishes at a value 
$\tau_{\Omega}^*\sim 6,000\Delta t$. Although we do not yet have a 
complete understanding of the mechanism of the current reversal, we 
believe that $\tau_{\Omega}^*$ is related to the intra-well relaxation 
time in equilibrium, which is estimated to be $5,236\Delta t$ (see the inset 
of the lower graphs of Fig. \ref{fig:b1}).   In Ref.\cite{hanggi2}, it 
is asserted that the intra-well relaxation time  determines the 
oscillation period, which causes the current reversal in a rocking 
ratchet system.  Taking these into consideration, the current reversal 
observed  in our protein model seems to be similar to that phenomenon 
originally reported in Ref. \cite{hanggi2}.

\section{Discussion}

In conclusion, we investigated motion of  F$_1$-ATPase in the absence of 
ATP by performing numerical simulations of the coarse-grained protein model, 
and fabricated a rocking ratchet applying the ratchet theory studied 
in Ref. \cite{hanggi2} to  F$_1$-ATPase.  We observed the current reversal 
when we changed the oscillation frequency and predicted the directions of 
rotation when there exists an oscillating force.  We hope that our 
predictions for a rocking ratchet consisting of F$_1$-ATPase in a 
nucleotide-free state will be confirmed experimentally, and that 
combination of theories of non-equilibrium statistical mechanics with 
protein simulations will yield new understanding of statistical properties 
possessed by proteins \cite{hayataka,ike}.  Below, we discussed two things 
related to our results.  

It should be noted that the parameter values used in our simulation do 
not correspond to room temperature and the real time scale of experiments.  
The temperature used in our simulations corresponds to one third less than 
the room temperature  in order for the $\gamma$-subunit not to lose its 
shape when the oscillating force is applied. However, we believe that 
the basic prediction of our model - that the direction of rotation can 
be reversed by altering the period of the oscillating force - would be 
unchanged if such parameter values were used.  In previous numerical 
studies \cite{koga,pu} also, temperatures much less than room temperature 
were used in order to make the coupling between  the structure change 
of the $\beta$-subunits and the rotation of the $\gamma$-subunit 
``tight,''  on the basis of the idea of the push-pull model \cite{wang}. 
However, it is worth simulating the motion of F$_1$-ATPase with 
realistic parameter values that represent the physiological conditions 
in which the molecular softness and thermal fluctuations would be more 
important.   With such conditions in mind, the ratchet-based mechanism 
has been considered instead of the push-pull model \cite{sakuraba} in 
an attempt to account for the results of a recent experiment 
\cite{furuike}.  The same idea has been applied to an F$_{\rm o}$-motor 
system \cite{sakuraba}.  

%

One reason for investigating systems like that considered here is to 
gain an understanding of motor proteins that could lead to the design 
and fabrication of new nano-machines that can operate efficiently in 
thermal noise. For this purpose, it is important to be able to cause 
directional motion of proteins not through use of ATP hydrolysis but by 
other energy sources. For this reason, we have studied a system in which 
the fluctuations and softness of F$_1$-ATPase are exploited in such a 
manner to cause rotation through the influence of an  oscillating force.  
Because the ratchet theories \cite{prost,hanggi1,hanggi2} are regarded 
as providing a theoretical foundation for the design of nano-machines,  
we hope that this study provides a basis for the development of one class 
of such machines.

\section{Appendix} Here we describe the model studied in this work. 
The position and velocity of the $i$th $\alpha$-carbon atom in F$_1$-ATPase 
are denoted by $\vecr_i$ and $\vecv_i$.  The Hamiltonian is given by 
$H(\{\vecr_i,\vecv_i\})=\sum_{i=1}^{N+n} m|\vecv_i|^2/2+V_{\alpha_3\beta_3}
+V_{\gamma}+V_{\rm int}+V_{\rm fix}$, where $m$ is the mass of the $i$th 
$\alpha$-carbon atom and $N$ and $n$ are the number of $\alpha$-carbon atoms 
in the $\alpha_3\beta_3$-subunits and in the $\gamma$-subunit, respectively.  
The quantity $V_{\alpha_3\beta_3}$ is the elastic network potential for 
the $\alpha_3\beta_3$ subunits and is given by 
\begin{eqnarray}
&&V_{\alpha_3\beta_3}(\{\vecr_i\}_{\alpha_3\beta_3})=\sum_{i=1}^{N-1}
\frac{k}{2}(|\vecr_i-\vecr_{i+1} | -|\vecr_i^0-\vecr_{i+1}^0|)^2 \nonumber \\ 
&+&\sum_{i=1}^{N-2}
\frac{0.5k}{2}(|\vecr_i-\vecr_{i+2} | -|\vecr_i^0-\vecr_{i+2}^0|)^2    
\nonumber \\
&+&\sum_{i=1}^{N-1}\sum_{j=i+3}^{N}
\frac{0.1k}{2}(|\vecr_i-\vecr_{j} | -|\vecr_i^0-\vecr_{j}^0|)^2,  
\label{eq:enm}
\end{eqnarray}
where $\{\vecr^0 \}$ represents the reference structure, and  we have $k=0$ 
when $|\vecr_i^0-\vecr_j^0| > r_{\rm c}$.  The elastic network potential,  
$V_{\gamma}$, is defined similarly to $V_{\alpha_3\beta_3}$.  Here,  
$\{\vecr^0 \}_{\alpha_3\beta_3}$ is composed of   the 
$\alpha_{\rm E}\beta_{\rm E}$-subunits of the crystal structure 1BMF 
\cite{walker}, which was taken from the RSCB Protein DataBank \cite{pdb}. 
We rotate them  by $\pm 120^\circ$ in the $xy$-plane and choose the reference 
structure as 
$\{ \alpha_{\rm E}\beta_{\rm E}, \alpha_{\rm E}\beta_{\rm E}, \alpha_{\rm E}
\beta_{\rm E} \}$, where the $xy$-plane is defined so that the 
three $\beta_9$ are on the plane. The set $\{\vecr^0 \}_{\gamma}$ is also 
constructed by using  1BMF. The interaction potential between the 
$\gamma$-subunit and the $\alpha_3\beta_3$-subunits consists of a 
repulsive potential and an electrostatic potential: 
\begin{eqnarray}
V_{\rm int}(\{\vecr_i\})  
&=&\sum_{i=1}^{N}\sum_{j=N+1}^{N+n} \left\{ 
\varepsilon_1\left(\frac{D}{|\vecr_i-\vecr_j|}\right)^{12} \right. 
\nonumber \\ 
&+& \left. \varepsilon_2 \frac{q_iq_j}{|\vecr_i-\vecr_j|^2} \right\}. 
\label{eq:pote}
\end{eqnarray}
Here, 
$q_i$ represents the sign of the electric charge of the amino  acid around 
the $i$th $\alpha$-carbon atom: $q_i=1$ when the amino acid is lysine 
or arginine,  $q_i=-1$ when it is glutamic acid or aspartic acid, 
and $q_i=0$ otherwise.  In our model, $\gamma_{272}$ and the three $\beta_9$ 
are fixed by including the potential $V_{\rm fix}(\vecr_i)=k_{\rm h}
|\vecr_i-\vecr_i^0|^2/2$. 

The time evolution of the $i$th $\alpha$-carbon atom is described by an 
over-damped Langevin equation with friction coefficient $\Gamma$ and 
 temperature $T$. In our simulation, the parameters are set to 
be $m=1$, $k=10$,  $N=2823$, $n=122$, $r_{\rm c}=10$, $D=6$, $\Gamma=0.01$, 
$r_{\rm c}=10$, $T=0.3$,  $\varepsilon_1=3.0$, $\varepsilon_2=0.05$ and 
$k_{\rm h}=0.5$.  We chose these values so that the results obtained 
without the oscillating force are consistent with the results observed 
experimentally \cite{fujisawa}. 

Regarding the definition of the rotation angle $\theta(t)$, as is done 
in the previous study \cite{koga}, we consider the projection 
$\veclam_{xy}(t)$  of $\veclam(t)=\sum_{i\in \gamma79-90} [\vecr_i(t)-
\vecr_{\gamma 78}(t)]/ |\vecr_i(t)-\vecr_{\gamma 78}(t)|$ onto the $xy$-plane. 
$\theta(t)$ is defined by the angle between $\veclam_{xy}(t)$ and 
$\veclam_{xy}^{\rm 1BMF}$ where $\veclam_{xy}^{\rm 1BMF}$ is the 
corresponding vector calculated for 1BMF\cite{walker}. 

\section{Acknowledgment}

We thank R. Fujisawa, R. Iino and H. Noji for discussions on  
experimental results for F$_1$-ATPase in a nucleotide-free state.  One of 
the authors (K. H.) thanks S. Toyabe, E. Muneyuki, R. Kanada and F. Takagi 
for discussions at the beginning of this work, and S. Toyabe for informing 
her of Ref. \cite{hanggi2}. This work is supported by JSPS fellowship 
(to K. H.) and Grand-in Aids for Scientific Research (no. 19042020 to 
H. Y. and M. T.) from MEXT in Japan.

\end{document}